\documentclass[11pt,a4paper]{article}
\usepackage{graphicx,amssymb,amsmath}
\usepackage{epsfig}
\usepackage{epsf}
\usepackage{times}
\usepackage{amsfonts}
\usepackage{latexsym}
\usepackage{bbm}
\usepackage{here}

\newtheorem{theorem}{Theorem}

\newtheorem{corollary}{Corollary}

\newtheorem{conjecture}{Conjecture}

\newcommand{\later}[1]{{}}
\newcommand{\old}[1]{{}}
\long\def\ignore#1{}

\newcommand{\RR}{\mathbb{R}} 

\providecommand{\intd}[0]%
{\;\mbox{d}}

\title{On stars and Steiner stars. II}

\author{Adrian Dumitrescu\thanks{Department of Computer Science,
University of Wisconsin-Milwaukee,  WI 53201-0784, USA. E-mail: {\tt
ad@cs.uwm.edu}}
\and Csaba D. T\'oth\thanks{Department of Mathematics,
University of Calgary, AB, Canada T2N~1N4. E-mail: {\tt
cdtoth@ucalgary.ca}}
\and
Guangwu Xu\thanks{Department of Computer Science,
University of Wisconsin-Milwaukee, WI 53201-0784, USA, email: {\tt
gxu4uwm@uwm.edu}}}

\begin{document}

\maketitle

\begin{abstract}
A {\em Steiner star} for a set $P$ of $n$ points in $\RR^d$ connects an
arbitrary  center point to all points of $P$, while a
{\em star} connects a point $p\in P$ to the remaining $n-1$ points of
$P$. All connections are realized by straight line segments.
Fekete and Meijer showed that the minimum star is at most $\sqrt{2}$ times
longer than the minimum Steiner star for any finite point configuration in
$\RR^d$. The maximum ratio between them, over all finite point configurations
in $\RR^d$, is called the {\em star Steiner ratio} in $\RR^d$.
It is conjectured that this ratio is $4/\pi = 1.2732\ldots$ in the plane and
$4/3=1.3333\ldots$ in three dimensions. Here we give upper bounds of $1.3631$
in the plane, and $1.3833$ in 3-space, thereby substantially improving recent
upper bounds of $1.3999$, and $\sqrt{2}-10^{-4}$, respectively.
Our results also imply improved bounds on the maximum ratios between the
minimum star and the maximum matching in two and three dimensions.

Our method exploits the connection with the classical problem of
estimating the maximum sum of pairwise distances among $n$ points on
the unit sphere, first studied by L\'aszl\'o Fejes T\'oth. It is quite general
and yields the first non-trivial estimates  below $\sqrt2$ on the star Steiner
ratios in arbitrary dimensions.  We show, however, that the star Steiner ratio
in $\RR^d$ tends to $\sqrt{2}$, the upper bound given by Fekete and Meijer,
as $d$ goes to  infinity. Our estimates on the  star Steiner ratios are
therefore much closer to the conjectured values in higher dimensions!
As it turns out, our estimates as well as the conjectured values of the
Steiner ratios (in the limit, for $n$ going to infinity) are related to the
classical infinite Wallis product:
$ \frac{\pi}{2}= \prod_{n=1}^{\infty} \left(\frac{4n^2}{4n^2-1}\right)
= \frac21 \cdot \frac23 \cdot \frac43 \cdot \frac45
\cdot \frac65 \cdot \frac67 \cdot \frac87 \cdot \frac89 \cdots $.
\end{abstract}

\section{Introduction} \label{sec:intro}

The study of minimum Steiner stars and minimum stars is motivated by
applications in facility location and  computational statistics
\cite{BMS99,BMM03,CT90,E00}.  The {\em Weber point},
also known as the {\em Fermat-Toricelli point} or {\em Euclidean
median}, is the point of the space that minimizes the sum of distances
to $n$ given points in $\RR^d$. The problem of finding such a point can
be asked in any metric space.
It is known that even in the plane, the Weber point cannot
be computed exactly, already for $n\geq 5$~\cite{B88,CM69}.
(For $n=3$ and 4, resp., Torricelli and Fagnano gave algebraic solutions.)
The Weber center can however be approximated with arbitrary
precision~\cite{BMM03,CT90}, mostly based on Weiszfeld's
algorithm~\cite{W37}. The reader can find more information on this
problem in ~\cite{DKSW02}, and in the recent paper of the
the first two named authors~\cite{DT08}.

The maximum ratio between the lengths of the minimum star and the
minimum Steiner star, over all finite point configurations in $\RR^d$,
is called the {\em star Steiner ratio} in $\RR^d$, denoted by $\rho_d$.
The same ratio for a specific value of $n$ is denoted by $\rho_d(n)$.
Obviously $\rho_d(n) \leq \rho_d$, for each $n$.
Fekete and Meijer~\cite{FM00} were the first to study the star Steiner
ratio. They proved that $\rho_d \leq \sqrt{2}$ holds for any dimension $d$.
It is conjectured that $\rho_2=4/\pi= 1.2732\ldots$, and $\rho_3 =
4/3=1.3333\ldots$, which are the limit ratios for a uniform mass distribution
on a circle,  and surface of a sphere, respectively
(in these cases the Weber center is the center of the circle, or
sphere)~\cite{FM00}.
By exploiting these bounds, Fekete and Meijer also established bounds
on the maximum ratio $\eta_d$, between the length of the minimum star
and that of a maximum matching on a set of $n$ points ($n$ even) in
two and three dimensions ($d=2,3$).

In a recent paper, the upper bounds on the Steiner ratios for $d=2,3$
have been lowered to $1.3999$ in the plane, and to $\sqrt{2}-10^{-4}$
in 3-space \cite{DT08}. The method of proof used there was not
entirely satisfying, as it involved heavy use of linear
programming. Besides that, the proofs were quite involved and the
improvements were rather small, particularly in three-space.
It was also shown in \cite{DT08} that the bound $4/\pi$ holds in two
special cases, corresponding to the lower bound construction; details
at the end of Section \ref{sec:plane}.

In this paper we get closer to the core of the problem, obtain
substantially better upper bounds, and moreover replace the use of linear
programming by precise and much shorter mathematical proofs.
Here we prove that $\rho_2\leq 1.3631$, and $\rho_3 \leq 1.3833$.
Based on these estimates, we can then further improve the estimates
given on $\eta_2$ and $\eta_3$, using the method developed
by Fekete and Meijer~\cite{FM00}.
Our improvements are summarized in Table \ref{table}.
Here $\min{S}$ denotes a minimum star and its length (with a slight
abuse of notation), $SS^*$ denotes a minimum Steiner star and its
length, and $\max{M}$ denotes a maximum length matching and its
length (for an even number of points).
Finally, our method yields the first non-trivial estimates  below $\sqrt2$ on
the star Steiner ratios in arbitrary dimensions. Among others, we
show that the upper bound $\sqrt2$ on the star Steiner ratio given
Fekete and Meijer is in fact a very good approximation of this ratio
for higher dimensions $d$; thus in this sense the problem in the
plane is the most interesting one.

\medskip
\begin{table}[hbtp]
\begin{center}
\begin{tabular}{||c|c|c|c||} \hline
Ratio & Lower bound & Old upper bound & New upper bound \\ \hline
$ \rho_2: (\min{S})/SS^*$ & $\frac{4}{\pi}=1.2732\ldots$ & $1.3999$
& $1.3631$ \ \ $\dagger$ \\ \hline
$\rho_3: (\min{S})/SS^* $ & $\frac{4}{3}=1.3333\ldots$ &
$\sqrt{2}-10^{-4}=1.4141\ldots$ & $1.3833$  \ \ $\dagger$ \\ \hline
$\rho_4: (\min{S})/SS^* $ & $\frac{64}{15 \pi}=1.3581\ldots$ \ \ $\dagger$ &
$\sqrt{2}=1.4142\ldots$ & $1.3923$  \ \ $\dagger$ \\ \hline
$\rho_5: (\min{S})/SS^* $ & $\frac{48}{35}=1.3714\ldots$ \ \ $\dagger$ &
$\sqrt{2}=1.4142\ldots$ & $1.3973$  \ \ $\dagger$ \\ \hline
$\rho_{100}: (\min{S})/SS^* $ & $1.4124\ldots$ \ \ $\dagger$ &
$\sqrt{2}=1.4142\ldots$ & $1.4135$  \ \ $\dagger$ \\ \hline \hline
$ \eta_2: \min{S} / \max{M}$ & $\frac{4}{3}=1.3333\ldots$ & $1.6165$ $$
& $1.5739$  \ \ $\dagger$ \\ \hline
$\eta_3: \min{S} / \max{M}$ & $\frac{3}{2}=1.5$ & $1.9999 $
& $1.9562 $  \ \ $\dagger$ \\ \hline
\end{tabular}
\end{center}
\caption{\small Lower and upper bounds on
star Steiner ratios ($\rho_2, \rho_3, \rho_4, \rho_5, \rho_{100}$),
and matching ratios ($\eta_2, \eta_3$) for some small values of $d$. Those
marked with $\dagger$ are new.}
\label{table}
\end{table}

If $q_0, q_1, \ldots, q_{n-1}$ are $n$ variable points on the unit (radius)
sphere in $\RR^d$, let $G(d,n)$ denote the maximum value of the
function $\sum_{i<j} | q_i q_j|$, i.e., the the maximum value of the
sum of pairwise distances among the points.
It was shown by Fejes T\'oth \cite{FT56}, and rediscovered in
\cite{DT08}, that $G(2,n)$ has a nice expression in closed form:
\begin{equation} \label{E1}
G(2,n)= \frac{n}{\tan{\frac{\pi}{2n}}}
= \frac{2}{\pi} n^2 - \frac{\pi}{6} + O\left( \frac{1}{n^2} \right).
\end{equation}

However, only the simpler inequality $G(2,n) \leq \frac{2}{\pi} n^2$ will be
needed here.
The {\em exact} determination of $G(d,n)$ for $d \geq 3$ is considered to be
a difficult geometric discrepancy problem \cite[pp. 298]{ABC04},
however estimates of the  form $G(d,n) \leq c_d n^2/2$ are known~\cite{Bj56},
where $c_d$ is the ``constant of uniform distribution'' for the sphere
\cite{A72,ABC04,AS74}: $c_d$ equals the average inter-point distance
for a uniform mass distribution on the surface of the unit sphere in $\RR^d$.
In particular for $d=3$, Alexander \cite{A72,A77} has shown that
\begin{equation} \label{E2}
\frac{2}{3} n^2 - 10 n^{1/2} < G(3,n) < \frac{2}{3} n^2 - \frac{1}{2}.
\end{equation}

Here $c_2=4/\pi$, and $c_3=4/3$. The connection with this problem
is explained in the next section.

\section{Stars in the plane} \label{sec:plane}

Fix an arbitrary coordinate system.
Let $P=\{p_0,\ldots,p_{n-1}\}$ be a set of $n$ points in the Euclidean plane,
and let $p_i=(x_i, y_i)$, for $i=0,\ldots,n-1$.
Let $SS^*$ be a minimal Steiner star for $P$, and assume that its
center $c=(x,y)$ is {\em not} an element of $P$.
As noted in ~\cite{FM00}, the minimality of the Steiner star implies
that the sum of the unit vectors rooted at $c$ and oriented to the
points vanishes, i.e.,
$\sum_{i=0}^{n-1} \frac{\overrightarrow{cp_i}}{|\overrightarrow{cp_i}|}=0$.
For completeness, we include here the brief argument (omitted in ~\cite{FM00}).
The length of the star centered at an arbitrary point $(x,y)$ is
$$ L(x,y)= \sum_{i=0}^{n-1} \sqrt{(x-x_i)^2 + (y-y_i)^2}. $$
If $(x,y)$ is the Weber center, we have
$$ \frac{\partial}{\partial x} L(x,y) = \frac{\partial}{\partial y}
L(x,y) = 0. $$
The two equations give
\begin{equation} \label{E3}
\sum_{i=0}^{n-1} \frac{x-x_i}{\sqrt{(x-x_i)^2 + (y-y_i)^2}} = 0 \hspace{5mm}
{\rm and} \hspace{5mm}
\sum_{i=0}^{n-1} \frac{y-y_i}{\sqrt{(x-x_i)^2 + (y-y_i)^2}} = 0.
\end{equation}

\begin{figure}[htbp]
\centerline{\epsfxsize=5.5in \epsffile{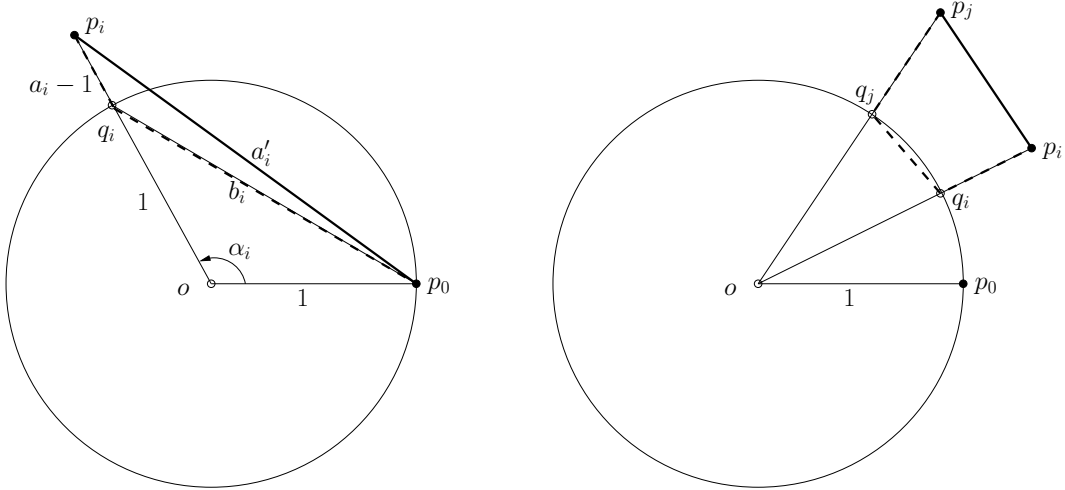}}
\caption{\small Left: estimating the length of a star centered at
$p_0=(1,0) \in P$. Right: estimating pairwise distances.}
\label{f1}
\end{figure}

Our setup is as follows. Refer to Figure \ref{f1}.
We may assume w.l.o.g.  that the Weber center is the origin $o=(0,0)$.
We may also assume that the Weber center is not in $P$, since otherwise
the ratio is $1$.
We can assume w.l.o.g.  that the closest point in $P$ to $o$ is
$p_0=(1,0)$, hence $SS^*=(1+\delta)n$, for some $\delta \geq 0$.
Let $C$ be the unit radius circle centered at $o$.
We denote by $\overrightarrow{v}$ a vector $v$,
and by $|\overrightarrow{v}|$ its length.
Write $\overrightarrow{r_i}=\overrightarrow{o p_i}$, for
$i=0,1,\ldots,n-1$, and let $\overrightarrow{o q_i}
=\frac{\overrightarrow{o p_i}}{|\overrightarrow{o p_i}|}$
be the corresponding unit vector; i.e., $q_i$ is the intersection
between $\overrightarrow{r_i}$ and the the unit circle $C$.
Let $a_i=|\overrightarrow{r_i}|$, $b_i=|\overrightarrow{p_0 q_i}|$,
and $a'_i=|\overrightarrow{p_0 p_i}|$, for $i=0,1,\ldots,n-1$.
We have $SS^*=\sum_{i=0}^{n-1} a_i$.
Finally, let $\alpha_i \in [0,2\pi)$ be the angle between the positive $x$-axis
and $\overrightarrow{r_i}$.

Henceforth \eqref{E3} can be rewritten in the more convenient form
\begin{equation}\label{E4}
\sum_{i=0}^{n-1} \cos \alpha_i=0 \hspace{5mm}
{\rm and} \hspace{5mm}
\sum_{i=0}^{n-1} \sin \alpha_i=0.
\end{equation}
Since the orientation of the coordinate system was chosen arbitrarily, such
formulas hold for any other orientation.
Note that this is also equivalent with sum of unit vectors vanishing:
$\sum_{i=0}^{n-1} \overrightarrow{o q_i}=0$.
Moreover, such formulas hold for any dimension $d \geq 2$.

Let $S_i$ be the star (and its length) centered at $p_i$, for
$i=0,1,\ldots,n-1$, and let  $\min{S}=\min_{0 \leq i \leq n-1} S_i$
denote the minimum star.
Using the local optimality condition \eqref{E4},
Fekete and Meijer~\cite{FM00} show that if one moves the center of
$SS^*$ from the Weber center to a closest point of $P$, the sum of
distances increases by a factor of at most $\sqrt{2}$; this bound which is
best possible for this method implies that for any $d \geq 2$ (see
~\cite{FM00} for details):
\begin{equation} \label{E5}
\rho_d(n) \leq \sqrt2, \ \textrm{ thus } \ \rho_d \leq \sqrt2.
\end{equation}
From the opposite direction, by considering a uniform mass distribution on the
surface of a unit sphere in $\RR^d$, one has for any $d \geq 2$:
\begin{equation} \label{E6}
\rho_d \geq c_d.
\end{equation}

Our new argument is a nutshell is as follows. If $\delta$ is large,
we consider the star centered at a point in $P$ closest to the Weber
center, as a good candidate for approximating the minimum star.
If $\delta$ is small, we use an averaging argument
to upper bound the length of the minimum star. In the end we balance
the two estimates obtained.
Applying the averaging argument (for small $\delta$) leads naturally
to the problem of maximizing the sum of pairwise distances among $n$
points on the surface of the unit sphere (or unit circle).

\begin{theorem}\label{T1}
The star Steiner ratio in the plane is less than $1.3631$.
More precisely:
$$ \frac{4}{\pi} \leq \rho_2 \leq
\frac{2\sqrt2 - \frac{4}{\pi}}{1+\sqrt2 - \frac{4}{\pi}} < 1.3631. $$
\end{theorem}
\begin{proof}
Consider an $n$-element point set $P$, and the previous setup.
It is enough to show that
$$ \frac{\min{S}}{SS^*} \leq
\frac{2\sqrt2 - \frac{4}{\pi}}{1+\sqrt2 - \frac{4}{\pi}} \ . $$
By the triangle inequality (see also Fig. \ref{f1}(left)~), we have
$a'_i \leq b_i+a_i-1$, for $i=0,\ldots,n-1$. Hence
$$ S_0= \sum_{i=0}^{n-1} a'_i \leq \sum_{i=0}^{n-1} (b_i + a_i -1) =
 \sum_{i=0}^{n-1} b_i + \delta n. $$
By Lemma 4 in \cite{FM00}, the local optimality condition \eqref{E4}
implies $\sum_{i=0}^{n-1} b_i \leq n\sqrt{2}$.  It follows then that
$ S_0 \leq (\sqrt2 +\delta)n $. Hence the star Steiner ratio
is at most
\begin{equation} \label {E7}
\frac{S_0}{SS^*} \leq \frac{\sqrt2 +\delta}{1 + \delta}.
\end{equation}

Since the local optimality condition $\sum_{i=0}^{n-1} \cos \alpha_i=0$
holds for any $d$, \eqref{E7} also holds for any $d$.
Observe that
$$ \lim_{\delta \rightarrow 0} \frac{\sqrt2 +\delta}{1 + \delta}=
\sqrt2, \textrm{ and }
\lim_{\delta \rightarrow \infty} \frac{\sqrt2 +\delta}{1 + \delta}=1, $$
with the above expression being a decreasing function of $\delta$
for $\delta \geq 0$.

Clearly, the sum of the lengths of the  $n$ stars (centered at each of the $n$
points) equals twice the sum of pairwise distances among the points.
$$ \sum_{i=0}^{n-1} S_i = 2 \sum_{i<j} | p_i p_j|. $$
By the triangle inequality (see also Fig. \ref{f1}(right)~)
$$ | p_i p_j| \leq | p_i q_i| + | q_i q_j| + | q_j p_j| =
(a_i -1) + | q_i q_j| + (a_j -1). $$
By summing up over all pairs $i<j$, we get
$$ \sum_{i=0}^{n-1} S_i \leq 2 \sum_{i<j} | q_i q_j| +
2(n-1)\sum_{i=0}^{n-1} (a_i -1) =
2 \sum_{i<j} | q_i q_j| + 2 \delta (n-1)n. $$

Using the upper bound estimate on $G(2,n)$ in \eqref{E1} we obtain
$$ \sum_{i=0}^{n-1} S_i \leq \frac{2n}{\tan{\frac{\pi}{2n}}}
+ 2 \delta (n-1)n \leq \frac{4}{\pi} \hspace{2pt} n^2 + 2 \delta n^2. $$
The minimum of the $n$ stars, $\min{S}$,  clearly satisfies:
$$ \min{S} \leq \frac{\sum_{i=0}^{n-1} S_i}{n}
\leq \left(\frac{4}{\pi} \hspace{2pt} + 2 \delta\right) n. $$

It follows that the star Steiner ratio is at most
\begin{equation} \label {E8}
\frac{\min{S}}{SS^*} \leq \frac{\frac{4}{\pi} + 2\delta}{1+\delta}.
\end{equation}
This estimate holds for any dimension $d$: the points $q_i$ lie on the surface
of the unit radius sphere $B$ in $\RR^d$ centered at $o$ (rather than
on the unit circle $C$).  Observe that
$$ \lim_{\delta \rightarrow 0} \frac{\frac{4}{\pi} + 2\delta}{1+\delta}
= \frac{4}{\pi}, \textrm{ and }
\lim_{\delta \rightarrow \infty} \frac{\frac{4}{\pi} + 2\delta}{1+\delta}
=2, $$
with the above expression being an increasing function of $\delta$
for $\delta \geq 0$.
Therefore, by combining this observation with the previous
observation following \eqref{E7}, we get
$$ \frac{\min{S}}{SS^*} \leq \max_{\delta \geq 0}
\min \left(\frac{\sqrt2 +\delta}{1 + \delta},
\frac{\frac{4}{\pi} + 2\delta}{1+\delta} \right). $$
The maximum value is given by substituting for $\delta$ the
solution of the equation
$ \sqrt2 +\delta = \frac{4}{\pi} + 2\delta $ (that is, by balancing
the two upper estimates on the Steiner ratio given by inequalities
\eqref{E7} and \eqref{E8}~).
The solution $\delta_0= \sqrt2 - \frac{4}{\pi} = 0.1409\ldots$
yields
\begin{equation} \label{E9}
\rho_2(n) \leq
\frac{2 \sqrt2 - \frac{4}{\pi}}{1 + \sqrt2 -  \frac{4}{\pi}}= 1.3630\ldots,
\ \textrm{ thus also } \
\rho_2 \leq \frac{2 \sqrt2 - \frac{4}{\pi}}{1 + \sqrt2 -  \frac{4}{\pi}}=
1.3630\ldots.
\end{equation}
%
\end{proof}

By the result in \cite{FM00}, $SS^* \leq \frac{2}{\sqrt{3}} \max{M}$.
By our Theorem \ref{T1},
$\min{S} \leq \frac{2\sqrt2 -\frac{4}{\pi}}{1+\sqrt2 - \frac{4}{\pi}} \cdot
SS^*$. Combining the two upper bounds yields the following upper bound on
$\eta_2$:

\begin{corollary}\label{C1}
The minimum star to maximum matching ratio in the plane ($\eta_2$) is
less than $1.5739$. That is, for any point set
$$ \frac{\min{S}}{\max{M}} \leq
\frac{2\sqrt2 - \frac{4}{\pi}}{1+\sqrt2 - \frac{4}{\pi}} \cdot
\frac{2}{\sqrt3} \leq 1.5739. $$
\end{corollary}
The best known lower bound for this ratio, is $4/3$, see \cite{FM00}.

\medskip
According to \cite[Theorem 2]{DT08},
the star Steiner ratio for a set of $n$ points in the plane that lie
on a circle centered at the Weber center is at most
$$ \frac{\frac{\pi}{2n}}{\tan{\frac{\pi}{2n}}} \cdot \frac{4}{\pi} <
\frac{4}{\pi}. $$
The same bound holds for any finite point set in the plane where the
angles from the Weber center to the $n$ points are uniformly distributed
(that is, $\alpha_i=2i\pi/n$, for $i=0,1,\ldots,n-1$)~\cite[Theorem 3]{DT08}.
We think that this is always the case, and thereby venture a slightly
stronger version of the conjecture proposed by Fekete and Meijer~\cite{FM00}
(however, this is for specific values of $n$):
\begin{conjecture}\label{J1}
The star Steiner ratio for $n$ points in the plane is
$$ \rho_2(n) =
\frac{\frac{\pi}{2n}}{\tan{\frac{\pi}{2n}}} \cdot \frac{4}{\pi}. $$
\end{conjecture}

\section{Stars in the space} \label{sec:space}

In this section we give estimates on the star Steiner ratio in $3$-space
(Theorem \ref{T2}), and in higher dimensions (Theorem \ref{T3}).

\begin{theorem}\label{T2}
The star Steiner ratio in $\RR^3$ is less than $1.3833$.
More precisely:
$$ \frac{4}{3} \leq \rho_3 \leq \frac{2}{17} (16-3\sqrt2)<1.3833. $$
\end{theorem}
\begin{proof}
The method of proof and the setup is the same as in the planar case,
so we omit the details.
Let $B$ be the unit radius sphere centered at $o$, analogous to
the unit circle $C$. Now all the points $q_i$ lie on the surface
of $B$. Using the upper bound estimate on $G(3,n)$ in \eqref{E2} we get the
analogue of Equation \eqref{E8}:
\begin{equation} \label {E10}
\frac{\min{S} }{SS^*} \leq
\frac{\frac{4}{3} + 2\delta}{1+\delta}.
\end{equation}
Taking also \eqref{E7} into account, we have
$$ \rho_3 \leq \max_{\delta \geq 0}
\min \left(\frac{\sqrt2 +\delta}{1 + \delta},
\frac{\frac{4}{3} + 2\delta}{1+\delta} \right). $$
By balancing the two upper estimates in \eqref{E7} and \eqref{E10}
as in the planar case yields
$ \delta_0= \sqrt2 - \frac{4}{3} = 0.0808\ldots$, and
\begin{equation} \label{E13}
\rho_3 \leq \frac{2 \sqrt2 - \frac{4}{3}}{1 + \sqrt2 -  \frac{4}{3}}=
\frac{6\sqrt2 -4}{3\sqrt2 -1}= \frac{2}{17} (16-3\sqrt2) =
1.3832\ldots.
\end{equation}
\end{proof}

By the result in \cite{FM00}, $SS^* \leq \sqrt{2} \cdot \max{M}$.
By our Theorem \ref{T2},
$\min{S} \leq \frac{2}{17} (16-3\sqrt2) \cdot SS^*$.
Combining the two yields the following upper bound on $\eta_3$:

\begin{corollary}\label{C2}
The minimum star to maximum matching ratio in $3$-space ($\eta_3$) is
less than $1.9562$. That is, for any point set
$$ \frac{\min{S}}{\max{M}} \leq \frac{2}{17} (16-3\sqrt2) \sqrt2 =
\frac{4}{17} (8 \sqrt2 - 3)  <1.9562. $$
\end{corollary}
The best known lower bound for this ratio, is $3/2$, see \cite{FM00}.

\medskip
The same method we used in proving Theorem \ref{T1} and
Theorem \ref{T2}, together with various approximations
yield the following estimates on the star Steiner ratio in $\RR^d$:

\begin{theorem}\label{T3}
Let $1<c_d<2$ be the ``constant of uniform distribution'' for the sphere
in $\RR^d$, $d \geq 2$. The star Steiner ratio in $\RR^d$ is bounded as
follows:
$$ c_d \leq \rho_d \leq \frac{2\sqrt2 - c_d}{1+\sqrt2 -c_d}, \textrm{ where }
\lim_{d \rightarrow \infty} c_d =
\lim_{d \rightarrow \infty} \rho_d =\sqrt2. $$
The following closed formula approximations hold:
$$ \sqrt2 e^{-\frac{1}{4(2d-3)}} \leq \rho_d \leq
\frac{2\sqrt2 - \sqrt2 e^{-\frac{1}{5(2d-1)}}}
{1+\sqrt2 - \sqrt2 e^{-\frac{1}{5(2d-1)}}} . $$
\end{theorem}
\begin{proof}
Note that by the same argument used in the proofs of Theorem \ref{T1} and
Theorem \ref{T2} (equations \eqref{E9} and \eqref{E13}), we have
\begin{equation} \label {E11}
c_d \leq \rho_d \leq \frac{2\sqrt2 - c_d}{1+\sqrt2 -c_d}.
\end{equation}

In order to establish the limits, we start by computing the ``constant of
uniform distribution'' $c_d$.
Recall that $c_d$ equals the average distance from a point on the
unit sphere in $\RR^d$ to all the other points on the same sphere,
for a uniform mass distribution.
It is easy to verify that $c_d$ is given by the following integral formula:
\begin{equation} \label{E15}
c_d = \frac{2
\int_{0}^{\pi/2} \sin^{d-2} (2\alpha) \cdot \sin \alpha \intd \alpha}
{\int_{0}^{\pi/2} \sin^{d-2} (2\alpha) \intd \alpha}.
\end{equation}
Some initial values are
$$ c_1 = 1, \ \
c_2 = \frac{\pi}{4}=1.2732\ldots, \
c_3 = \frac{4}{3}=1.3333\ldots, \
c_4 = \frac{64}{15\pi}=1.3581\ldots, \
c_5 = \frac{48}{35}=1.3714\ldots. $$

In order to establish a recurrence on $c_d$, define
$$ a_{ij}= \int_{0}^{\pi/2} \sin^i \alpha  \cdot \cos^j \alpha \intd \alpha ,
\ \ i,j \geq 0. $$
Some initial values are
$$ a_{00}=\frac{\pi}{2}, \ \ a_{01}=a_{10}=1, \ \
a_{11}=\frac{1}{2}, \ \ a_{02}=a_{20}=\frac{\pi}{4}. $$

Expanding $\sin 2\alpha$ yields then
$$ c_d = \frac{2 a_{d-1,d-2}}{a_{d-2,d-2}}. $$
Recall that integration by parts leads to the well-known
recurrence relations for $a_{ij}$, for $i,j \geq 1$:
\begin{eqnarray*}
a_{ij} = \int_{0}^{\pi/2} \sin^i \alpha  \cdot \cos^j \alpha \intd \alpha &=&
-\frac{\sin^{i-1} \alpha  \cdot \cos^{j+1} \alpha }{i+j}  \Big{|}_{0}^{\pi/2} +
\frac{i-1}{i+j} \int_{0}^{\pi/2} \sin^{i-2} \alpha  \cdot \cos^j \alpha
\intd \alpha\\
&=& \frac{\sin^{i+1} \alpha \cdot \cos^{j-1} \alpha }{i+j}
\Big{|}_{0}^{\pi/2} +
\frac{j-1}{i+j} \int_{0}^{\pi/2} \sin^i \alpha  \cdot \cos^{j-2} \alpha
\intd \alpha.
\end{eqnarray*}
Plugging these in the formula for $c_d$ immediately gives a recurrence
for $c_d$. For any $d \geq 1$:
$$ c_{d+2}=
\frac{2 \cdot \frac{d}{2d+1} \cdot \frac{d-1}{2d-1} \cdot a_{d-1,d-2}}
{\frac{d-1}{2d} \cdot \frac{d-1}{2d-2} \cdot a_{d-2,d-2}}=
\frac{4d^2}{4d^2-1} c_d = \left( 1+ \frac{1}{4d^2-1} \right) c_d . $$

Recall at this point the infinite Wallis product from number
theory\cite{HW79}:
$$ \frac{\pi}{2}= \prod_{k=1}^{\infty} \left(\frac{4k^2}{4k^2-1}\right)
= \frac21 \cdot \frac23 \cdot \frac43 \cdot \frac45
\cdot \frac65 \cdot \frac67 \cdot \frac87 \cdot \frac89 \cdots . $$
Let
$$ W_n = \prod_{k=1}^n \left(\frac{4k^2}{4k^2-1}\right),
\ \ \textrm{ and } \ \
Z_n = \prod_{k=n+1}^{\infty} \left(\frac{4k^2}{4k^2-1}\right), $$
denote the partial finite and respectively partial infinite
Wallis products, so that
$ W_n Z_n =\pi/2$, for every $n \geq 1$.
Our recurrence for $c_d$ yields
that $c_d$ is an increasing sequence satisfying also
\begin{equation} \label{E12}
c_{d+1} c_{d+2} = c_1 c_2 W_d , \ \ d \geq 1.
\end{equation}
Since $c_d$ is bounded, it converges to some limit $c$. The value of $c$
can be obtained by solving the equation
$$ c^2 = c_1 c_2 \frac{\pi}{2} =2. $$
We thus have $\lim_{d \rightarrow \infty} c_d =\sqrt2$.
Since $c_d \leq \rho_d \leq \sqrt2$, we also have
$\lim_{d \rightarrow \infty} \rho_d =\sqrt2$.
From Equation \eqref{E12}, we also get that for $d \geq 3$
\begin{equation} \label{E14}
c_1 c_2 W_{d-2} \leq c_d^2 \leq c_1 c_2 W_{d-1}, \ \ \textrm{ or }\ \
\sqrt{c_1 c_2 W_{d-2} } \leq c_d \leq \sqrt{c_1 c_2 W_{d-1}}.
\end{equation}

Observe that
$$ \sum_{k=n+1}^{\infty} \frac{1}{4k^2-1}=
\frac{1}{2} \sum_{k=n+1}^{\infty} \left(\frac{1}{2k-1}- \frac{1}{2k+1} \right)
=\frac{1}{2(2n+1)}. $$
Standard inequalities\footnote{Here we have chosen $4x/5$ for simplicity
of resulting expressions.}
$ e^{4x/5} \leq 1+x \leq e^x$ for $x \in [0,1/3]$
now imply that for each $n \geq 1$
$$ Z_n = \prod_{k=n+1}^{\infty} \left(1+\frac{1}{4k^2-1}\right)
\leq e^{\sum_{k=n+1}^{\infty} \frac{1}{4k^2-1}}=e^{\frac{1}{2(2n+1)}}, $$
and
$$ Z_n = \prod_{k=n+1}^{\infty} \left(1+\frac{1}{4k^2-1}\right)
\geq e^{\frac{4}{5} \sum_{k=n+1}^{\infty}
\frac{1}{4k^2-1}}=e^{\frac{2}{5(2n+1)}}. $$
Since $W_n = (\pi/2)/Z_n$, we have
$$ \frac{\pi}{2} \cdot e^{-\frac{1}{2(2n+1)}} \leq
W_n \leq \frac{\pi}{2} \cdot e^{-\frac{2}{5(2n+1)}}, $$
and consequently \eqref{E14} gives
$$ \frac{2}{\sqrt{\pi}} \cdot \frac {\sqrt{\pi}} {\sqrt2}
\cdot e^{-\frac{1}{4(2d-3)}} \leq
c_d \leq \frac{2}{\sqrt{\pi}} \cdot \frac {\sqrt{\pi}} {\sqrt2}
\cdot e^{-\frac{1}{5(2d-1)}}, $$
or equivalently
$$ \sqrt2 e ^{-\frac{1}{4(2d-3)}} \leq c_d \leq
\sqrt2 e ^{-\frac{1}{5(2d-1)}}. $$
Taking into account \eqref{E11} and subsituting the above upper bound on
$c_d$, we finally get the estimate (for any $d \geq 4$):
$$ \sqrt2 e^{-\frac{1}{4(2d-3)}} \leq \rho_d \leq
\frac{2\sqrt2 - \sqrt2 e^{-\frac{1}{5(2d-1)}}}
{1+\sqrt2 - \sqrt2 e^{-\frac{1}{5(2d-1)}}} . $$

\vspace{-1.33\baselineskip}
\end{proof}

For the values of $c_d$ given by \eqref{E15}, we can extend the
conjecture of Fekete and Meijer to all dimensions $d \geq 2$:

\begin{conjecture}\label{J2}
The star Steiner ratio in $\RR^d$ equals the ``constant of uniform
distribution'' for the sphere in $\RR^d$: that is, $\rho_d=c_d$
\ for any $d \geq 2$.
\end{conjecture}

\end{document}